# Magnetic structure and phase diagram of TmB$_4$

S. GABÁNI, S. MAŤAŠ, P. PRIPUTEN, K. FLACHBART

Institute of Experimental Physics, Slovak Academy of Sciences, Watsonova 47, SK-04001 Košice, Slovakia

K. SIEMENSMEYER, E. WULF

Hahn Meitner Institut, Glienicker Str. 100, D-14109 Berlin, Germany

A. EVDOKIMOVA, N. SHITSEVALOVA

Institute for Problems of Materials Science, Ukrainian Academy of Sciences, Krzhyzhanovsky 3, UA-03680 Kiev, Ukraine

Magnetic structure of single crystalline TmB$_4$ has been studied by magnetization, magnetoresistivity and specific heat measurements. A complex phase diagram with different antiferromagnetic (AF) phases was observed below $T_{N1}$ = 11.7 K. Besides the plateau at half-saturated magnetization (1/2 $M_S$), also plateaus at 1/9, 1/8 and 1/7 of $M_S$ were observed as function of applied magnetic field $B//c$. From additional neutron scattering experiments on TmB$_4$, we suppose that those plateaus arise from a stripe structure which appears to be coherent domain boundaries between AF – ordered blocks of 7 or 9 lattice constants. The received results suggest that the frustration among the Tm$^{3+}$ magnetic ions, which maps to a geometrically frustrated Shastry-Sutherland lattice lead to strong competition between AF and ferromagnetic (FM) order. Thus, stripe structures in intermediate field appear to be the best way to minimize the magnetostatic energy against other magnetic interactions between the Tm – ions combined with very strong Ising anisotropy.

## 1. Introduction

Rare earth tetraborides REB$_4$ crystallize in a tetragonal structure with the space group *P4/mbm* where the RE ions map to a Shastry-Sutherland type geometrically frustrated lattice (SSL) in the *c*-plane. It was shown that all heavy REB$_4$ (RE = Tb, Dy, Ho, Er, Tm) exhibit strong Ising-like anisotropy which orients the RE magnetic moments along the *c*-axis, and complex phase diagrams below $T_N$(TbB$_4$) = 43 K, $T_N$(DyB$_4$) = 20.3 K, $T_N$(HoB$_4$) = 7.1 K, $T_N$(ErB$_4$) = 15.4 K and $T_N$(TmB$_4$) = 11.7 K [1-5]. Moreover, in the ordered AF state of TmB$_4$, the magnetization $M$ for $B//c$ reaches saturation $M_S$ at about 4 T accompanied by plateaus at 1/8 $M_S$ and 1/2 $M_S$. On the other hand, for $B\perp c$ the 1/2 $M_S$ state and the saturation of $M$ were observed only at much higher field above 30 T [6]. Such behavior may result from the combined effect of the crystal electric field (CEF) and the magnetic frustration attributed to the SSL formed by the Tm$^{3+}$ ions.

The aim of this work is to confirm and extend recent studies of the magnetic structure of TmB$_4$ by investigations of macroscopic physical quantities like magnetization, resistivity and specific heat.

## 2. Experimental details

A large single crystal of TmB$_4$ was prepared by a floating zone melting technique. The *c*-axis and *a*-axis of the crystal were determined by Laue photographs. Oriented small samples were cut parallel to these directions. The temperature and field dependence of magnetization, *M(T, B)*, up to 14 T, and in temperature range from 1.8 K to 300 K was measured using a vibrating sample magnetometer of Quantum Design (PPMS). DC resistivity measurements between 1.5 K and 300 K, $\rho(T, B)$, were carried out in a $^4$He flow cryostat with a 7 T magnet using the standard four-terminal method. Specific heat measurements down to 0.3 K and up to 5 T, *C(T, B)*, were made by the relaxation method in a $^3$He cryostat of Oxford Instruments (Kelvinox).

## 3. Results and discussion

TmB$_4$ shows strong anisotropy in the temperature dependence of the susceptibility $\chi(T)$ for $B//c$ and $B\perp c$ (see inset Fig. 1(a)). The Néel temperatures $T_{N1}$ = 11.7 K and $T_{N2}$ = 9.8 K are the same for both field directions, and the effective moment value is close to that of a free Tm$^{3+}$ ion, $\mu_{eff}(B//c)$ = 6.5 and $\mu_{eff}(B\perp c)$ = 6.7. However, the paramagnetic Curie temperature $\theta_P$ is positive, $\theta_P$ = 45.9 K, for $B//c$, whereas it is negative, $\theta_P$ = -57.6 K, for $B\perp c$. Most interesting is the *M(B)* dependence for $B//c$, where besides the plateau at half-saturated magnetization (1/2 $M_S$) in high field also plateaus of $M(B)/M_S$ = 1/9, 1/8 and 1/7 are observed in fields around $B$ = 1.6 T at low temperature (Fig. 1(a)). In addition, $\chi(T)_{B\perp c}$ has a broad maximum around 85 K which likely comes from the CEF splitting. For $B\perp c$, just above $T_{N1}$, magnetization plateaus are seen as a function of temperature. The plateau magnetization depends linearly on the applied field which suggests a quadratic Zeeman effect with a splitting of 14 meV between ground and first excited state.

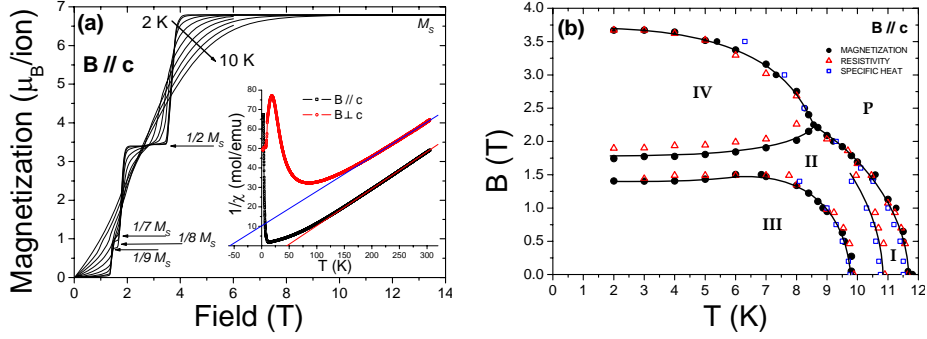

Fig.1. Magnetization data (a) and the final *B-T* phase diagram (b) of $TmB_4$ for *B//c*.

The magnetic phase diagram (see Fig. 1(b)) was constructed from the peak positions of numerically calculated *dχ(T)/dT* and *dM(B)/dB*, and shows tree phases (II, III and IV) below $T_{N1}$ with one triple point at $(T_t, B_t) \approx$ (8.6 K, 2.2 T).

Preliminary results of the neutron scattering experiments of $TmB_4$ reveal a simple AF structure of the uu-dd type in zero magnetic field. The ferrimagnetic high field phase (phase IV) seems to consist of large blocks with a size of 8 chemical unit cells. The data suggest that AF correlations between the dimers ($Tm^{3+}$-$Tm^{3+}$) are dominant in the *c*-plane and FM interactions along the *c*-axis. The properties of the intermediate phase (II) are more complex and show hysteresis effects both as a function of temperature and magnetic field. We suppose that the magnetization plateaus at $M(B)/M_S$ = 1/7 and 1/9 seen in intermediate field arise from a magnetic *stripe structure* as was suggested for $TbB_4$ in [2]. In essence, the stripes appear to be coherent domain boundaries between AF - ordered blocks of 7 or 9 lattice constants.

The magnetoresistivity measurements, *ρ(T, B)*, as shown on Fig. 2(a), detect the additional phase (phase I) between paramagnetic and ordered phase II. The value of *RRR* ≈ 114 indicates the high quality of the sample. The analysis by spin wave scattering formula $\rho(T) \propto T^x exp(-\Delta_{sw}/k_B T)$ below $T_{N2}$ gives the parameter $x \approx 2$, typical for FM ordering, and spin wave activation energy $\Delta_{sw} \approx$ 2.4 meV, which decreases down to ≈ 1 meV with increasing magnetic field.

The temperature dependence of the specific heat *C(T)* up to 5 T confirms the existence of phase I. In the paramagnetic state the magnetic part of specific heat, $C_{mag}(T, B)$, can very well be fitted by a Schottky anomaly with a parameter $T_{max} \approx$ 48 K and gap energy *ΔE* ≈ 9.9 meV. The magnetic entropy at zero field was calculated from $C_{mag}/T$ vs *T* dependence and is shown in Fig. 2(b). The total molar entropy below $T_{N1}$ is *Rln1.75* and the contribution below $T_{N2}$ being only *Rln1.42*, similar as for $HoB_4$ [1]. The site symmetry of the $Tm^{3+}$ ion is *mm* and the CEF can therefore lift all the degeneracy of the $^3H_6$ ground state of $Tm^{3+}$. The fact that the entropy of $TmB_4$ in ordered state is less than *Rln2* makes it possible that Tm$^{3+}$ has a ground state of two closely lying singlets, nevertheless, the magnetization data suggest a doublet ground state with $J_m = 6$.

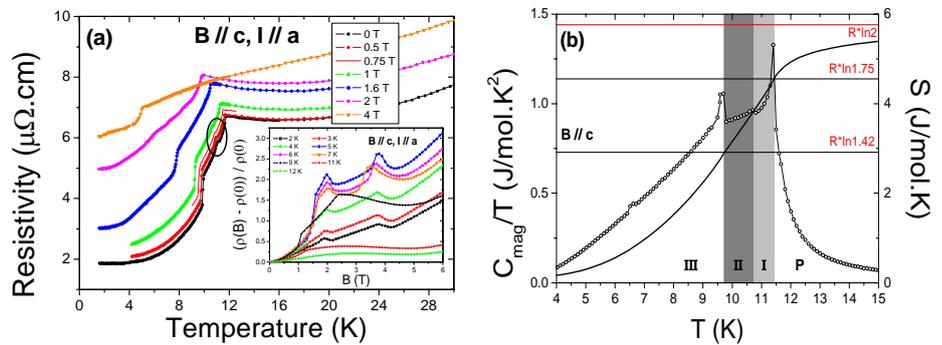

Fig.2. Magnetoresistivity (a) and magnetic specific heat (b) of TmB$_4$ for $B//c$.

In summary, TmB$_4$ exhibits a strong Ising-like anisotropy. The combination of the CEF anisotropy and of magnetic frustration leads to strong competition between AF and FM order. Consequently, complex magnetic structures are found in the various magnetic phases of TmB$_4$. The stripe structure in intermediate field appears to be the best way to minimize the magnetostatic energy against other magnetic interactions between Tm ions combined with the very strong Ising anisotropy. Interestingly, this leads to intermediate stripe phases based on odd integers (1/7, 1/9) when the sample history favours AF order, the even occurrence (1/8) is found for the opposite FM case.

It is a pleasure to thank K. Kiefer (HMI) for help with the PPMS magnetometer and the specific heat experiments. This work was supported by the DAAD, Slovak scientific agency VEGA – contract 7054, by the Project of the Slovak Research and Development Agency APVV 031704, and by the contract I/2/2003 of the Slovak Academy of Sciences for the Centre of Excellence. Liquid nitrogen for experiments has been sponsored by U.S. Steel Košice, s. r. o..